\documentclass[12pt,showpacs,preprintnumbers,superscriptaddress,amsmath,amssymb,nofootinbib]{revtex4}
\usepackage{graphicx}
\usepackage{dcolumn}
\usepackage{bm}
\usepackage{amssymb}
\usepackage{amsmath}
\usepackage{epsfig}    
\usepackage{color}
\usepackage{hhline}

\def\be{\begin{equation}}
\def\ee{\end{equation}}
\newcommand{\bea}{\begin{eqnarray}}
\newcommand{\eea}{\end{eqnarray}}
\newcommand{\nn}{\nonumber}

\numberwithin{equation}{section}

\begin{document}

\title{A  Radiative  Neutrino Model with  $SU(2)_L$ Triplet Fields}

\author{Takaaki Nomura}
\email{nomura@kias.re.kr}
\affiliation{School of Physics, KIAS, Seoul 02455, Korea}

\author{Hiroshi Okada}
\email{macokada3hiroshi@cts.nthu.edu.tw}
\affiliation{Physics Division, National Center for Theoretical Sciences, Hsinchu, Taiwan 300}

\author{Yuta Orikasa}
\email{orikasa@kias.re.kr}
\affiliation{School of Physics, KIAS, Seoul 02455, Korea}
\affiliation{Department of Physics and Astronomy, Seoul National University, Seoul 08826, Korea}
\affiliation{Institute of Experimental and Applied Physics, Czech Technical University in Prague,
128 00 Prague 2, Czech Republic}

\date{\today}

\begin{abstract}
We propose a loop induced neutrino mass model, in which  we introduce several exotic fermions and bosons with  $SU(2)_L$ multiplet, and discuss various phenomenologies such as lepton flavor violations,  muon anomalous magnetic moment, nonstandard interacting neutrinoless double beta decay, relic density of dark matter, and the possibility of the spin independent direct detection searches, imposing the constraints of  oblique parameters. And we show a benchmark point to satisfy all the constraints and discuss our predictions.
\end{abstract}
\maketitle
\newpage

\section{Introduction}
Radiatively induced neutrino mass models are one of the promising candidates to accommodate dark matter (DM) candidate, and several exotic fermions and/or bosons in a low energy scale($\sim$ TeV scale).
{Also these fields are often linked to the neutrino masses that gives several predictions and various phenomenologies.} 
Hence a lot of authors have historically been working along this ideas. Here we classify such radiative models as the number of the loops,
{\it i.e.}, ref.~\cite{a-zee, Cheng-Li, Pilaftsis:1991ug, Ma:2006km, Gu:2007ug, Sahu:2008aw, Gu:2008zf, AristizabalSierra:2006ri, Bouchand:2012dx, McDonald:2013hsa, Ma:2014cfa, Kajiyama:2013sza, Kanemura:2011vm, Kanemura:2011jj, Kanemura:2011mw, Schmidt:2012yg, Kanemura:2012rj, Farzan:2012sa, Kumericki:2012bf, Kumericki:2012bh, Ma:2012if, Gil:2012ya, Okada:2012np, Hehn:2012kz, Dev:2012sg, Kajiyama:2012xg, Toma:2013zsa, Kanemura:2013qva, Law:2013saa, Baek:2014qwa, Kanemura:2014rpa, Fraser:2014yha, Vicente:2014wga, Baek:2015mna, Merle:2015gea, Restrepo:2015ura, Merle:2015ica, Wang:2015saa, Ahn:2012cg, Ma:2012ez, Hernandez:2013dta, Ma:2014eka, Ma:2014yka, Ma:2015pma, Ma:2013mga, radlepton1, Okada:2014nsa, Brdar:2013iea, Okada:2015kkj, Bonnet:2012kz, Joaquim:2014gba, Davoudiasl:2014pya, Lindner:2014oea, Okada:2014nea, Mambrini:2015sia, Boucenna:2014zba, Ahriche:2016acx, Fraser:2015mhb, Fraser:2015zed, Adhikari:2015woo, Okada:2015vwh, Ibarra:2016dlb, Arbelaez:2016mhg, Ahriche:2016rgf, Lu:2016ucn, Kownacki:2016hpm, Ahriche:2016cio, Ahriche:2016ixu, Ma:2016nnn, Nomura:2016jnl, Hagedorn:2016dze, Antipin:2016awv, Nomura:2016emz, Gu:2016ghu, Guo:2016dzl, Hernandez:2015hrt, Megrelidze:2016fcs, Cheung:2016fjo, Seto:2016pks, Lu:2016dbc} mainly focusses on the scenarios at one-loop level, ref.~\cite{2-lp-zB, Babu:2002uu, AristizabalSierra:2006gb, Nebot:2007bc, Schmidt:2014zoa, Herrero-Garcia:2014hfa, Long:2014fja, VanVien:2014apa, Aoki:2010ib, Lindner:2011it, Baek:2012ub, Aoki:2013gzs, Kajiyama:2013zla, Kajiyama:2013rla, Baek:2013fsa, Okada:2014vla, Okada:2014qsa, Okada:2015nga, Geng:2015sza, Kashiwase:2015pra, Aoki:2014cja, Baek:2014awa, Okada:2015nca, Sierra:2014rxa, Nomura:2016rjf, Nomura:2016run, Bonilla:2016diq, Kohda:2012sr, Dasgupta:2013cwa, Nomura:2016ask, Nomura:2016pgg, Liu:2016mpf} at two-loop level,   ref.~\cite{Krauss:2002px, Aoki:2008av, Gustafsson:2012vj, Ahriche:2014xra, Ahriche:2014cda, Ahriche:2014oda, Okada:2014oda, Hatanaka:2014tba, Jin:2015cla, Culjak:2015qja, Okada:2015bxa, Geng:2015coa, Ahriche:2015wha, Nishiwaki:2015iqa, Okada:2015hia, Ahriche:2015loa, Kajiyama:2013lja, King:2014uha, Kanemura:2015bli, Okada:2016rav, Ko:2016sxg, Nomura:2016vxr, Thuc:2016qva, Cherigui:2016tbm, Nomura:2016ezz, Cheung:2016ypw, Cheung:2016frv, Gu:2016xno} at three-loop level,   
and ref.~\cite{Nomura:2016fzs, Nomura:2016seu} at four-loop level.

If the active neutrinos are Majorana type, it might be tested by searching for the neutrinoless double beta decay, which usually appears
on any types of canonical seesaw model via charged gauge boson in the standard model (SM). That is called standard interaction of neutrinoless double beta decay. However radiative neutrino models frequently include the non-standard interactions~\cite{Bergstrom:2011dt, Gustafsson:2014vpa} in addition to the standard one.
Thus it might be promising on the viewpoint of discriminating models, if experiments could show some anomalous results in the future.

In this paper, we introduce several exotic fermions and bosons with  $SU(2)_L$ multiplet
{including triplet exotic leptons and a {triplet Higgs} whose vacuum expectation value (VEV) is generated via trilinear scalar coupling induced at one-loop level.
In our model, small triplet VEV is naturally realized  which is required by the constraint from $\rho$-parameter.}
Then we discuss various phenomenologies {such as lepton flavor violating processes}
(especially $\mu-e$ conversions), muon anomalous magnetic moment, neutrinoless double beta decay from nonstandard interaction, {relic density and the possibility of the spin independent direct detection searches of DM}, imposing the constraints from oblique parameters. And we show a benchmark point to satisfy all the constraints and discuss our predictions.

This paper is organized as follows.
In Sec.~II, we introduce our model and show some formulae including Higgs masses, neutrino mass, LFV, muon anomalous magnetic  moment, and DM physics.
In Sec.~III, we show our numerical results. We conclude in Sec.~IV.

\section{The Model}


\begin{table}[thbp]
\centering {\fontsize{10}{12}
\begin{tabular}{|c||c|c|c|c|}
\hline Fermion & $L_L$ & $ e_{R} $ & $L'_{}$  & $\Sigma_{}$  
  \\\hhline{|=#=|=|=|=|$}
$(SU(2)_L,U(1)_Y)$ & $(\bm{2},-1/2)$ & $(\bm{1},-1)$ & $(\bm{2},-1/2)$   & $(\bm{3},-1)$
\\\hline
$U(1)_{H} $ & $0$ & $0$ &  $-x$ &  $-x$    \\\hline
$Z_2$ & $+$ & $+$ &  $-$&  $-$  \\\hline
\end{tabular}%
} \caption{Lepton sector; notice the three (or two) flavor index of each field $L_L$, $e_R$, $L'_{}$ and  $\Sigma_{}$ is abbreviated.} 
\label{tab:1}
\end{table}

\begin{table}[thbp]
\centering {\fontsize{10}{12}
\begin{tabular}{|c||c|c|c|c|c|}
\hline Boson  & $\Phi$   & $\eta$    & $\varphi$   & $\Delta$    & $\chi$ 
  \\\hhline{|=#=|=|=|=|=|}
$(SU(2)_L,U(1)_Y)$ & $(\bm{2},1/2)$  & $(\bm{2},1/2)$   & $(\bm{1},0)$   & $(\bm{3},1)$   & $(\bm{1},0)$ \\\hline
$U(1)_{H} $ & $0$ & $x$ &  $x$ & $2x$ &  $0$    \\\hline
$Z_2$ & $+$ & $-$ &  $+$ &  $+$ &  $-$  \\\hline
\end{tabular}%
} 
\caption{Boson sector }
\label{tab:2}
\end{table}

In this section, we introduce our model, in which the particle contents for leptons and bosons are respectively shown in Tab.~\ref{tab:1} and Tab.~\ref{tab:2}, and provide some formulae for such as neutrino mass matrix, lepton flavor violation processes and neutrino-less double beta decay.  
In our model, we add vector-like fermions  of $L'_{}$ with $SU(2)_L$ doublet and $\Sigma$ with $SU(2)_L$ triplet to the SM fields, where these fields have  $-x(\neq0)$ charge under the hidden local symmetry. Each of the exotic field needs (at least) two flavors in order to satisfy current neutrino oscillation data~\cite{pdg}. 
As for new bosons, we introduce
two neutral $SU(2)_L$ singlet scalars $\varphi$ and $\chi$  with  $U(1)_H$ charge $x$ and $0$ respectively,
an  $SU(2)_L$ doublet scalar $\eta$ with charge $x$, and an  $SU(2)_L$ triplet scalar $\Delta$ with charge $2x$.
We assume that $\Phi$, $\varphi$, and $\Delta$ have VEVs, which are symbolized by $v/\sqrt2$, $v'/\sqrt2$, and  $v_\Delta/\sqrt2$ respectively,
where VEVs of $\varphi$ and $\Delta$ spontaneously break  the hidden  symmetry down. 
Additional $Z_2$ symmetry plays a role in  assuring the stability of our DM candidate; the lightest mass eigenstate among neutral component of $L'_{}$, $\Sigma$, $\chi$ and $\eta$ which are $Z_2$ odd.

The relevant  renormalizable Lagrangian for Yukawa sector and scalar potential under these assignments
are given by
\begin{align}
-\mathcal{L}_{Y}
&=
(y_\ell)_i \bar L_{L_i} \Phi e_{R_i} + f_{ia} \bar L_{L_i} \Sigma_{R_a} \eta 
+ f'_{ab} \bar L'_{L_a}\Phi \Sigma_{R_b}
+ f''_{ab} \bar \Sigma _{L_a}\Phi^*  L'_{R_b}\nn\\
&+ (g_{R})_{ab} \bar L'^c_{R_a}(i\tau_2) \Delta L'_{R_b}
+(g_{L})_{ab} \bar L'^c_{L_a}(i\tau_2) \Delta L'_{L_b}
+(M_L)_{a}  \bar L'_{L_a} L'_{R_a} 
+(M_{\Sigma})_{a}  {\rm Tr[}\bar \Sigma_{L_a} \Sigma_{R_a}]  \nn\\
& -\mu_\chi (\chi^2 \varphi^* +{\rm h.c.}) -\mu_\eta (\eta^T(i\tau_2)\Delta^\dag \eta +{\rm h.c.})
-\lambda_0(\eta^\dag\Phi \chi\varphi+{\rm h.c.})
+\rm{h.c.} \label{Lag:Yukawa} 
\end{align}
where $\tau_2$ is the second Pauli matrix, each of the index $a(b)$ and $i(j)$ that runs $1$-$3$ represents the number of generations for exotic leptons and SM leptons respectively. 
We work on the basis where all the coefficients are real and positive for our brevity. 

{\it Scalar sector}:
After the electroweak symmetry breaking, each of scalar field has nonzero mass and
be parametrized  as~\cite{Okada:2015nca} 
\begin{align}
&\Phi =\left[
\begin{array}{c}
\phi^+\\
\phi^0
\end{array}\right],\
\eta =\left[
\begin{array}{c}
\eta^+\\
\eta^0
\end{array}\right],\
\Delta =\left[
\begin{array}{cc}
\frac{\Delta^+}{\sqrt2} & \Delta^{++}\\
\Delta^0 & -\frac{\Delta^+}{\sqrt2}
\end{array}\right],
\label{component}
\end{align}
where the neutral components of the above fields and the singlet scalar fields are defined by
\begin{align}
\phi^0&=\frac1{\sqrt2}(v + h + i \tilde a),\ \eta^0=\frac1{\sqrt2}(\eta_R+i \eta_I),\ 
 \Delta^0=\frac1{\sqrt2}(v_\Delta+\Delta_R+i \Delta_I),\nn\\
 \varphi &= \frac1{\sqrt2}(v'+\rho_R+i \rho_I), \quad \chi = \frac1{\sqrt2}(\chi_R+i \chi_I).
\label{Eq:neutral}
\end{align}
Here $v$ and $v_\Delta$ is related to the Fermi constant $G_F$ by $v^2+2 v^2_\Delta=1/(\sqrt{2}G_F)\approx(246$ GeV)$^2$.

{\it The CP even Higgs boson mass matrix with VEV}: $(M^{2})^{vev}_{\rm CP-even}$ in the basis of ($\Delta_R,h,\rho_R$) is diagonalized by 3 $\times$ 3 orthogonal mixing matrix $O_R$ as
$O_R (M^{2})^{vev}_{\rm CP-even} O_R^T=$ diag.$(m_{h_1}^2, m_{h_{\rm SM}}^2, m_{h_3}^2)$.
Here
 $h_2 = h_{\rm SM}$ is the SM Higgs and $h_1$ and $h_3$ are additional Higgs mass eigenstates. 
{Then $Z_2$- and CP-even components in Eq~(\ref{Eq:neutral}) are related to the mass eigenstates as follows:
 \begin{equation}
 \Delta_R = (O_R^T)_{1 A} h_A, \quad h = (O_R^T)_{2 A} h_A, \quad \rho_R = (O_R^T)_{3 A} h_A,
 \end{equation}
 where $A = 1-3$.
 }

{\it The CP odd Higgs boson mass matrix with VEV}: $(M^{2})^{vev}_{\rm CP-odd}$ in the basis of ($\Delta_I,\tilde a$) is diagonalized by 2 $\times$ 2 orthogonal mixing matrix $O_I$ as
$O_I (M^{2})^{vev}_{\rm CP-odd}  O_I^T=$ diag.$(0,m_{a}^2)$, where $m_{a}^2=\frac{\mu_{\rm eff}(v^2+4 v^2_\Delta)}{\sqrt2 v_\Delta}$ and
the massless mode $G_Z$ is absorbed by the neutral gauge boson $Z$ as a Nambu-Goldstone (NG) boson; $\mu_{\rm eff}(\sim \Phi^T(i\tau_2)\Delta^\dag\Phi)$ is induced at one-loop level as can be seen later.
{Then ($Z_2$-even)-(CP-odd) components in Eq~(\ref{Eq:neutral}) are related to the mass eigenstates $(a_1,a_2) = (a, G_Z)$ as follows:
 \begin{equation}
\Delta_I = (O_I^T)_{1 \alpha} a_\alpha, \quad \tilde a = (O_I^T)_{2 \alpha} a_\alpha,
 \end{equation}
 where $\alpha = 1,2$.}
{We note that mixing effect is negligibly small since the off-diagonal component of $(M^{2})^{vev}_{\rm CP-odd}$ is proportional to $v_\Delta$. 
Thus relations of $a_1 \simeq a$ and $a_2 \simeq G_Z$ are indicated.}

{\it The CP even inert Higgs boson mass matrix}: $(M^{2})^{inert}_{\rm CP-even}$ in the basis of ($\eta_R,\chi_R$) is diagonalized by 2 $\times$ 2 orthogonal mixing matrix $V_R$ as
$V_R (M^{2})^{inert}_{\rm CP-even}  V_R^T=$ diag.$(m_{H_1}^2,m_{H_2}^2)$.
{Then inert CP-even components in Eq~(\ref{Eq:neutral}) are related to the mass eigenstates as follows:
 \begin{equation}
\eta_R = (V_R^T)_{1 \alpha} H_\alpha, \quad \chi_R = (V_R^T)_{2 \alpha} H_\alpha.
 \end{equation}
 }

{\it The CP odd inert Higgs boson mass matrix}: $(M^{2})^{inert}_{\rm CP-odd}$ in the basis of ($\eta_I,\chi_I$) is diagonalized by 2 $\times$ 2 orthogonal mixing matrix $V_R$ as
$V_I (M^{2})^{inert}_{\rm CP-odd}  V_I^T=$ diag.$(m_{A_1}^2,m_{A_2}^2)$.
{Then inert CP-odd components in Eq~(\ref{Eq:neutral}) are related to the mass eigenstates as follows:
 \begin{equation}
\eta_I = (V_I^T)_{1 \alpha} A_\alpha, \quad \chi_I = (V_I^T)_{2 \alpha} A_\alpha.
 \end{equation}
 }

{\it The $Z_2$-even singly charged Higgs boson mass matrix}: $(M^{2})^{Z_2 \, \text{even}}_{\rm singly}$ in the basis of ($\Delta^+,\phi^+$) is diagonalized by 2 $\times$ 2 orthogonal mixing matrix $O_C$ as
$O_C (M^{2})^{Z_2 \, \text{even}}_{\rm singly} O_C^\dag=$ diag.$(0,m_{C}^2)$, where $m_{C}^2=\frac{(\sqrt2\mu_{\rm eff}-\lambda'_{\Phi\Delta}v_\Delta)(v^2+2 v^2_\Delta)}{2 v_\Delta}$ and
the massless mode $G_W$ is absorbed by the charged gauge boson $W^\pm$ as NG boson.
{Then $Z_2$ even singly charged components in Eq~(\ref{component}) are related to the mass eigenstates $(H_1^+, H_2^+(= G_W))$ as follows:
 \begin{equation}
\Delta^+ = (O_C^T)_{1 \alpha} H^+_\alpha, \quad \phi^+_R = (O_C^T)_{2 \alpha} H^+_\alpha.
 \end{equation}}
 {Notice that we have $\Delta^+ \simeq H^+_1$ and $\phi^+ \simeq G_W^+$ due to small off-diagonal component of mass matrix as in the $(M^{2})^{vev}_{\rm CP-odd}$.}
{\it The doubly charged boson mass matrix} does not have mixing. Thus its eigenvalue $m_{\Delta^{\pm \pm}}$ can be written in terms of a linear combinations of VEVs, trilinear term, and quartic couplings.

\begin{figure}[tbc]
\begin{center}
\includegraphics[width=70mm]{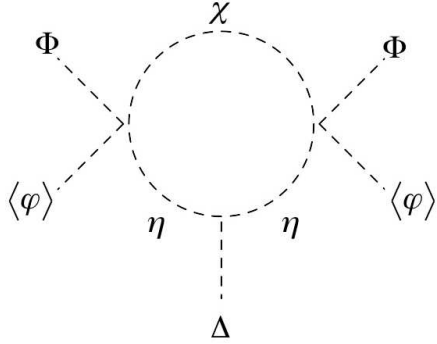}
\caption{The one loop diagram for generating effective trilinear coupling $\mu_{\rm eff}$.}
\label{loop1}
\end{center}
\end{figure}
{\it Effective trilinear coupling of  $\mu_{\rm eff}$}:
The coupling $\mu_{\rm eff}$, which is associated with the trilinear interaction $\Phi^T(i\tau_2)\Delta^\dag\Phi$ is induced not  at the leading order, but the one-loop level mediated by inert neutral bosons {as shown in Fig.~\ref{loop1}.}
 The effective trilinear coupling of  $\mu_{\rm eff}$ is given by
 \begin{align}
\mu_{\rm eff}&=\frac{(\lambda_0 v')^2\mu_\eta}{6(4\pi)^2}
\left[
({\cal V}_R)_{\alpha,\beta,\gamma}
{I}_3(m_{H_\alpha},m_{H_\beta},m_{H_\gamma})
-
({\cal V}_I)_{\alpha,\beta,\gamma}
{I}_3(m_{A_\alpha},m_{A_\beta},m_{A_\gamma})
\right],
 \end{align}
 with 
  \begin{align}
 ({\cal V}_R)_{\alpha,\beta,\gamma}\equiv {(V_R^T)_{2 \alpha}(V_R)_{\alpha 1} (V_R^T)_{1 \beta}(V_R)_{\beta 2} (V_R^T)_{1 \gamma}(V_R)_{\gamma 1}} ,\\
   ({\cal V}_I)_{\alpha,\beta,\gamma}\equiv {(V_I^T)_{2\alpha}(V_I)_{\alpha1} (V_I^T)_{1\beta}(V_I)_{\beta 2} (V_I^T)_{1 \gamma}(V_I)_{\gamma 1}}, \\
   {I}_3(m_1,m_2,m_3)\equiv\frac{1}{m_2^2-m_3^2}
   \left[\frac{m_2^2}{m_1^2-m_2^2}\ln\left[\frac{m_1^2}{m_2^2}\right]
   -
   \frac{m_3^2}{m_1^2-m_3^2}\ln\left[\frac{m_1^2}{m_3^2}\right]\right],
  \end{align}
where each of ($\alpha,\beta,\gamma$) runs form 1 to 2.
$v_\Delta$ is subsequently induced as $v_\Delta\sim \mu_{\rm eff}$~\cite{Kanemura:2012rj}.
Thus we have a theoretical reason that  $v_\Delta$ is tiny, which is in good agreement with the experimental result from the  $\rho$ parameter.

{\it Inert conditions}:
Since we have several charges bosons and negative contributions at the one-loop level, we have to take care in order not to have stable points of  their potentials{~\cite{Okada:2015nca}}. However since  these conditions are not so strong, we do not discuss in details. 

\subsection{Fermion Sector}
Let us fist define the exotic  fermion as follow:
\begin{align}
L'_{L(R)}\equiv 
\left[
\begin{array}{c}
N'\\
E'^-
\end{array}\right]_{L(R)},\quad
\Sigma_{L(R)}
\equiv
\left[\begin{array}{cc}
\Sigma^-/\sqrt2 & \Sigma_0 \\
\Sigma^{--} &  -\Sigma^-/\sqrt2 \\
\end{array}\right].
\label{eq:exotic_fermion}
\end{align}
{\it Neutral exotic fermion}:
Then the mass matrix for the neutral fermion in the basis of $\vec N_{L}\equiv [N'_L,N'^c_R,\Sigma_L,\Sigma_R^c]_{L}^T$ is given by 
\begin{align}
M_N=
\left[\begin{array}{cccc}
M_{N_L}  & M_L^* & 0 & M_{N\Sigma}^{'*} \\
M_L^\dag & M_{N_R}^\dag&  M_{N\Sigma}^{''\dag} &0 \\
0  &   M_{N\Sigma}^{''*} & 0 & M_\Sigma^* \\
M_{N\Sigma}^{'\dag} & 0 & M_\Sigma^\dag &0
\end{array}\right]
\sim
\left[\begin{array}{cccc}
M_{N_L}  & M_L^* & 0 & M_{N\Sigma}^{*} \\
M_L^* & M_{N_R}^*&  M_{N\Sigma}^{*} &0 \\
0  &   M_{N\Sigma}^{*} & 0 & M_\Sigma^* \\
M_{N\Sigma}^{*} & 0 & M_\Sigma^* &0
\end{array}\right],
\end{align}
where $M_{N_{L(R)}}\equiv g_{L(R)} v_\Delta/\sqrt2$, $M'_{N\Sigma}\equiv f'_{} v/\sqrt2$, $M''_{N\Sigma}\equiv f''_{} v/\sqrt2$, and we assume {three} generations case with positive real couplings with $f'\simeq f''$ for our simple analysis.
$M_N$ is diagonalized by unitary mixing matrix $V_N$ as
$V_N M_N V_N^T=M_N^{\rm diag.}$, and the mass eigenvector $\vec\psi_{L}\equiv [\psi_{1L},\psi_{2R}^c,\psi_{3_L},\psi_{4_R}^c]_{L}^T$ is defined by
$\vec N_{L}\equiv V_N^T \vec\psi_{L}$. Moreover, we also assume $M_N$ to be symmetric matrix, therefore, $M'_{N\Sigma}=M_{N\Sigma}^{'\dag}$, $M^{''}_{N\Sigma}=M_{N\Sigma}^{''\dag}$, $M_{\Sigma}=M_{\Sigma}^{\dag}$, and $M_{N_L}=M_{N_L}^{\dag}$.

{\it Singly charged exotic fermion}:
The mass matrix for the singly charged fermion in the basis of $\vec E\equiv [E'^-, \Sigma^-]_R^T$ is given by 
\begin{align}
M_E=
\left[\begin{array}{cc}
M_L & M'_{N\Sigma} \\
M_{N\Sigma}^{''} &  M_\Sigma \\
\end{array}\right]\sim
\left[\begin{array}{cc}
M_L & M_{N\Sigma} \\
M_{N\Sigma} &  M_\Sigma \\
\end{array}\right],
\end{align}
where we extract the one generation, and assume $M_{N\Sigma}\equiv M'_{N\Sigma}=M_{N\Sigma}^{''}$ for simplicity. Then the singly charged mass matrix can also be regarded as a symmetric one.
{Note here that the above assumption affects the neutral fermion mass matrix.}
Then $M_E$ is diagonalized by 6 $\times$ 6 unitary mixing matrix $V_C$ as $V_C M_E V_C^T=M_E^{\rm diag.}$, and the mass eigenvector $E^\pm$ is defined by $\vec E \equiv V_C^T \vec \psi^\pm $.

{In summary, exotic neutral and charged fermions in Eq.~(\ref{eq:exotic_fermion}) are written in terms mass eigenstates such that 
\begin{align}
& E'^\pm_a =  (V_C^T)_{a \kappa} \psi^\pm_\kappa, \, \, \Sigma^\pm_a = (V_C^T)_{3+ a, \kappa} \psi^\pm_\kappa  \\
& N'_{a L}  = (V_N^T)_{a n} \psi_{nL}, \, \, (N'_{a R})^c  = (V_N^T)_{3+a,n} \psi_{nL}, \, \, \Sigma^0_{a L} = (V_N^T)_{6+a,n} \psi_{nL}, \, \, \Sigma^{0c}_{a R} = (V_N^T)_{9+a,n} \psi_{nL}, \nn
\end{align}
where $\kappa = 1-6$ and $n = 1-12$.} {Note that the masses of doubly charged fermions $\Sigma^{\pm \pm}_a$ are given by $M_{\Sigma_a}$.}

\subsection{Electroweak precision test}

The $S$- and $T$-parameter constrain the masses and couplings of additional $SU(2)_L$ multiplet scalars and fermions.
Thus in our numerical analysis later, we impose constraint by the contributions to $S(T)$-parameters from new particles~{\cite{Khandker:2012zu}}.
The $S(T)$-parameters are calculated from vacuum polarization diagram for $Z$ and $W^\pm$ bosons, $i \Pi_{Z(W)}^{\mu \nu}$, where new particles run inside loop diagrams.
Then we obtain 
\begin{align}
\label{eq:piZ}
\Pi_{Z}^{\mu \nu}{(q^2)} &=  g^{\mu \nu} \frac{e^2}{c_W^2 s_W^2} \left( \Pi_{33}(q^2) - 2 s_W^2 \Pi_{3Q}{(q^2)}  - s_W^4 \Pi_{QQ}{(q^2)}  \right), \\
\label{eq:piW}
\Pi_{W}^{\mu \nu} &= g^{\mu \nu} \frac{e^2}{s_W^2} \Pi_{\pm}(q^2), \\
\Delta S & = \frac{4 e^2}{\alpha} \left[ \frac{d}{d q^2}\Pi_{33} (0) - \frac{d}{d q^2}\Pi_{3Q} (0) \right], \\ 
\Delta T & = \frac{ e^2}{\alpha s_W^2 c_W^2 m_Z^2} \left[\Pi_{\pm} (0) - \Pi_{33} (0) \right]. 
\end{align}
The list of new particle contributions to $\Pi_{33,3Q,QQ,\pm}$ is lengthy and we summarize them in the Appendix A.
Then we impose the constraint on the $S(T)$-parameters~\cite{pdg}:
\begin{align}
\Delta S &= 0.00 \pm 0.08, \\
\Delta T &= 0.05 \pm 0.07,
\end{align}
where we apply the constraint with $U$-parameter to be fixed as zero.
\footnote{{An interesting discussion is found in Refs.~\cite{Basso:2013jka, Antusch:2014woa, Akhmedov:2013hec}, which discuss the relation between the oblique parameter and the lepton universality. In our case, a new contribution to the lepton universality occurs only through the $f$ term at the box type of one-loop diagram, which could be subdominant. }}


\begin{figure}[tbc]
\begin{center}
\includegraphics[width=70mm]{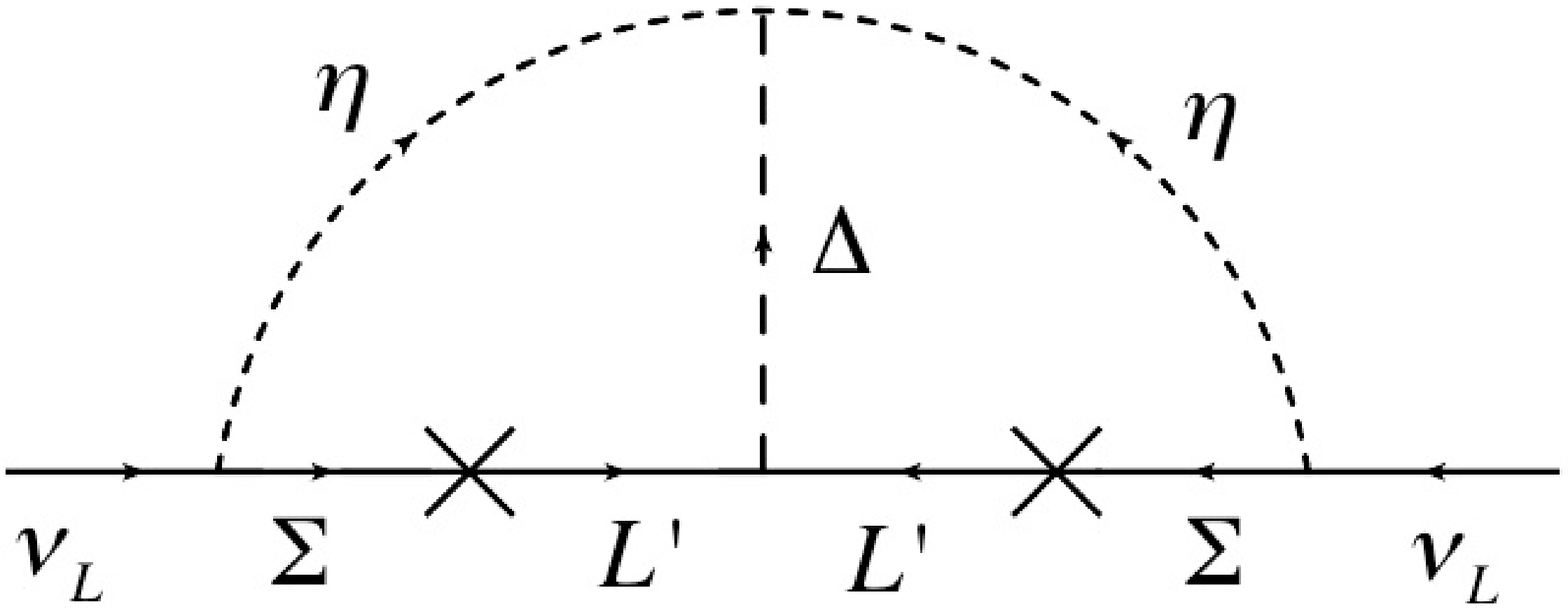}
\includegraphics[width=70mm]{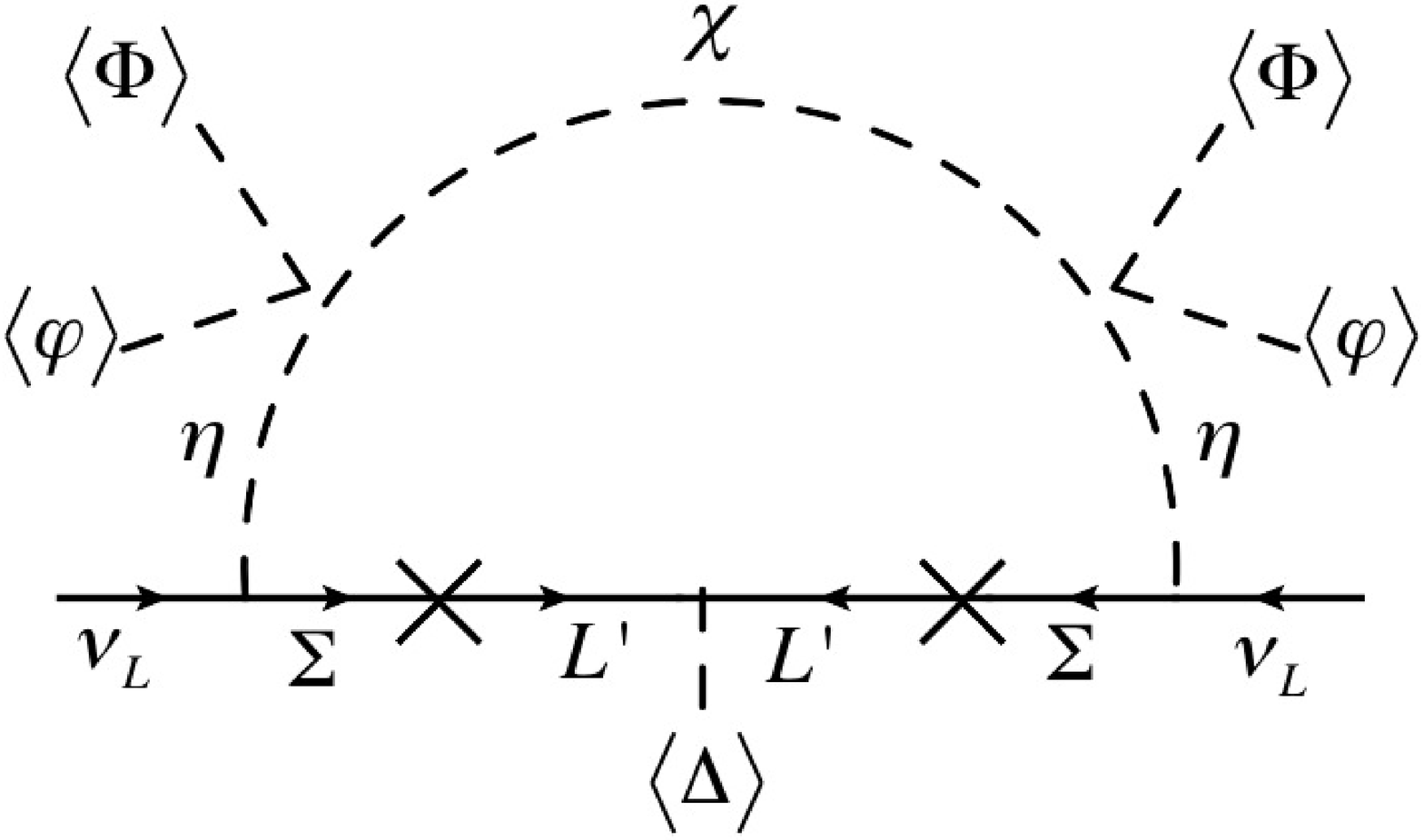}
\caption{The loop diagrams for generating active neutrino mass matrix where cross mark in the diagrams indicate $\langle \Phi \rangle$.}
\label{loop2}
\end{center}
\end{figure}
\subsection{Neutrino mass matrix}
{The neutrino mass matrix can be given by the linear combination at one-loop and  two-loop diagrams which are depicted in Fig.~\ref{loop2}, }
and their forms are respectively given by
\begin{align}
&({\cal M}_\nu)_{ij} = ({ m}_\nu)^{I}_{ij} +  ({ m}_\nu)^{II}_{ij},\\
&({m}_\nu)^{I}_{ij} 
=
{\frac{6 \delta m_{\chi^0 \eta^0}^2 \delta m_{\chi^0}^2}{(4\pi)^2} }
\sum_a^{1-3}(f)_{i,a} (M_{N\Sigma} {\delta M_L} M_{N\Sigma})_{a}  (f)_{j,a}\nn\\
&
\times\int \frac{dxdydzdw\delta(x+y+z+w-1)xyzw}{[x m^2_{\eta_0} +y m^2_{\chi_0} +z M^2_{N\Sigma_a}  + w M^2_{{N_L}_a}]^4}
\left[
3-\frac{4 M_{{N_L}_a}^2}{x m^2_{\eta_0} +y m^2_{\chi_0} +z M^2_{N\Sigma_a} + w M^2_{{N_L}_a}}\right],\\
&({ m}_\nu)^{II}_{ij} 
=
\frac{1}{(4\pi)^4}\frac{\mu_\eta}{v_\Delta}
\sum_a^{1-3}(f)_{i,a} (M_{N\Sigma} {\delta M_L} M_{N\Sigma})_{a}  (f)_{j,a}
\int \frac{dxdydzdw\delta(x+y+z+w-1)xyzw}{(w^2-w)(1-w)}\nn\\
&
\times\int \frac{dX dY dZ dW \delta(X+Y+Z+W-1)}
{-X \Delta_1 + Y M^2_{L_a} + Z M^2_{\Sigma_a} +W m^2_{\eta^\pm} }
\left[
2+w+\frac{M^2_{L_a}  }{-X \Delta_1 +Y M^2_{L_a} +Z M^2_{\Sigma_a} +W m^2_{\eta^\pm} }\right],\\
&
\Delta_1=
\frac{x M^2_{\Sigma_a} + y M^2_{L_a} +z m^2_{\eta^\pm}+w m^2_{\Delta^{\pm\pm}} }{w^2-w},
\end{align}
where we has used a mass insertion approximation method, {\it i.e.}, $M_{N_L(R)}, M_{N\Sigma}<<M_L,M_\Sigma$, {we has defined $\delta M_L \equiv g_L v_\Delta/\sqrt{2}$, $\delta m_{\chi^0 \eta^0}^2 \equiv \lambda_0 v v'/4$ and $\delta m_{\chi^0}^2 \equiv \mu_\chi v'/\sqrt{2}$, and $m_{\chi^0}$ and $m_{\eta^0}(=m_{\eta^\pm})$ denote mass parameter for quadratic terms of $\chi$ and $\eta$ before symmetry breaking respectively}. Hence Casas-Ibarra parametrization can be applicable~\cite{Casas:2001sr} because two of the loop functions are common and diagonal. Then it is convenient to decompose above  loop functions and compute separately, as shown in Appendix B.

Remind here that three flavors $\ell=1-3$ are introduced to obtain the current neutrino oscillation data~\footnote{{To minimally satisfy the neutrino oscillation data, two flavors are enough.}}.
Then $(\mathcal{M}_\nu)_{ab}$ can generally be diagonalized by the Maki-Nakagawa-Sakata mixing matrix $V_{\rm MNS}$ (MNS) as
\begin{align}
(\mathcal{M}_\nu)_{ab} &=(V_{\rm MNS} D_\nu V_{\rm MNS}^T)_{ab},\quad D_\nu\equiv (m_{\nu_1},m_{\nu_2},m_{\nu_3}),
\\
V_{\rm MNS}&=
\left[\begin{array}{ccc} {c_{13}}c_{12} &c_{13}s_{12} & s_{13} e^{-i\delta}\\
 -c_{23}s_{12}-s_{23}s_{13}c_{12}e^{i\delta} & c_{23}c_{12}-s_{23}s_{13}s_{12}e^{i\delta} & s_{23}c_{13}\\
  s_{23}s_{12}-c_{23}s_{13}c_{12}e^{i\delta} & -s_{23}c_{12}-c_{23}s_{13}s_{12}e^{i\delta} & c_{23}c_{13}\\
  \end{array}
\right],
\end{align}
where we neglect the Majorana phase as well as Dirac phase $\delta$ in the numerical analysis for simplicity.
We have used the following experimental values with best fit~\cite{Forero:2014bxa} 
\begin{eqnarray}
&&   s_{12}^2 \approx 0.306, \; 
 s_{23}^2 \approx0.42, \;
s_{13}^2 \approx0.021,  
  \\
&& 
  m_{\nu_1}= 0,  \; 
  \ m_{\nu_3}^2 =2.4 \times10^{-3} \ {\rm eV}^2,  \; 
  \ m_{\nu_2}^2 =(7.05) \times10^{-5} \ {\rm eV}^2, \nn
  \label{eq:neut-exp}
  \end{eqnarray}
where we assume one of three neutrino masses is zero with normal ordering in our analysis below.

\subsection{Neutrinoless double beta decay}
{
A non-standard contribution to the neutrinoless double beta decay is generated at the one-loop level,
and its form is given by~\cite{Gustafsson:2014vpa} 
}
\begin{align}
{\cal L}&=\frac{G_F^2}{2} m_p^{-1}(\epsilon_3)_{ij} J^\mu J_\mu j_{ij}+{\rm h.c.},\\
(\epsilon_3)_{ij}&= \epsilon_{R_{ij}}^{(1)}(H_1, V_{R_{12}})
+\epsilon_{L_{ij}}^{(1)}(H_1, V_{R_{12}})
+ \epsilon_{R_{ij}}^{(1)}(H_2, V_{R_{11}})
+\epsilon_{L_{ij}}^{(1)}(H_2, V_{R_{11}})
\nn\\
&+ \epsilon_{R_{ij}}^{(1)}(A_1, iV_{I_{12}})
+\epsilon_{L_{ij}}^{(1)}(A_1, iV_{I_{12}})
+ \epsilon_{R_{ij}}^{(1)}(A_2, iV_{I_{11}})
+\epsilon_{L_{ij}}^{(1)}(A_2, iV_{I_{11}}),
\end{align}
{where $m_p$ is the proton mass, $J^\mu=\bar u\gamma^\mu d$, $j_{ij}\equiv 2\bar \ell_i \ell^c_j$.}
Here {each of $\epsilon_3$ is found as
\begin{align}
\epsilon_{L_{ij}}^{(1)}(H_1,V_{R_{12}}) &\equiv
\sum_{n}^{1-12}\sum_{a,b}^{1-3}\sum_{\kappa,\zeta}^{1-6}\frac{f_{i a} f_{j b} (V_C^T)_{3+a,\kappa}(V_C)_{\kappa,3+a}
(V_C^T)_{3+b,\zeta}(V_C)_{\zeta,3+b}
 (V_N^T)_{6+a,n} (V_N^T)_{6+b,n} }{2^3(4\pi)^4}\nn\\
&\times \int\frac{dxdydzdw\delta(x+y+z+w-1)  V_{R_{12}}^2 M_{\psi^\pm_\kappa} M_{\psi_n} M_{\psi^\pm_\zeta}}
{[x M_{\psi^\pm_\kappa}^2+y M_{\psi_n}^2+z M_{\psi^\pm_\zeta}^2+w m_{H_1}^2]^2 },\\
\epsilon_{R_{ij}}^{(1)}(H_1, V_{R_{12}})&\equiv
\sum_{n}^{1-12}\sum_{a,b}^{1-3}\sum_{\kappa,\zeta}^{1-6}\frac{f_{i a} f_{j b} (V_C^T)_{3+a,\kappa}(V_C)_{\kappa,3+a}
(V_C^T)_{3+b,\zeta}(V_C)_{\zeta,3+b}
 (V_N^T)_{9+a,n} (V_N^T)_{9+b,k} }{2^3(4\pi)^4}\nn\\
&\times \int\frac{dxdydzdw\delta(x+y+z+w-1)  V_{R_{12}}^2 M_{\psi^\pm_\kappa} M_{\psi_n} M_{\psi^\pm_\zeta}}
{[xM_{\psi^\pm_\kappa}^2+yM_{\psi_n}^2+zM_{\psi^\pm_\zeta}^2+dm_{H_1}^2]^2 }.
\end{align}
}
This is known as the short distance effect, and $\epsilon_3\equiv (\epsilon_3)_{11(=ee)}$ can approximately be interpreted as 
\begin{align}
[T^{0\nu\beta\beta}_{1/2}]^{-1}\simeq G_{01}|\epsilon_3|^2 |{\cal M}^{SD}|^2,
\end{align}
where $G_{01}$ and ${\cal M}^{SD}$, which are dependent of the species of nuclei, are respectively given in Table 1 of Ref.~\cite{Gustafsson:2014vpa}.
For example, in case of $^{76}$Ge, the recent results, which is obtained by the phase I run of GERDA~\cite{Agostini:2013mzu} place, 
are given by $2.1\times 10^{25}\ {\rm yr} < T^{0\nu\beta\beta}_{1/2}$ at 90 \% confidential level (C.L.)
with $G_{01}\simeq 0.623\times 10^{-14}$ yr$^{-1}$ and ${\cal M}^{SD}\simeq 213$. Notice here that $^{76}$Ge currently gives the strongest bound on the neutrinoless double beta decay. As for the other bounds on nuclei, {\it i.g.}, we summarize Table~\ref{tab:nldbd} in Ref.~\cite{Gustafsson:2014vpa}.

\begin{table}[t]
\begin{tabular}{c|c|c|c} \hline
 & $G_{01}\ [10^{-14}$ yr$^{-1}$] & $|{\cal M}^\nu|$ & $|{\cal M}^{SD}|$ \\ \hline
$^{76}$Ge & 0.623 & 4.07 & 213 \\
$^{136}$Xe & 4.31 & 2.82 &109 \\
$^{150}$Nd  & 19.2 & 2.33 & 311 \\ 
$^{130}$Te  & 4.09 & 3.63 & 198 \\ 
$^{82}$Se  & 4270 & 3.48 & 192 \\ \hline
\end{tabular}
\caption{Summary of the experimental data of neutrinoless beta decay.}
\label{tab:nldbd}
\end{table}

\subsection{lepton flavor violations and muon anomalous magnetic moment}

\begin{table}[t]
\begin{tabular}{c|c|c} \hline
Process & $(i,j)$ & Experimental bounds ($90\%$ CL) \\ \hline
$\mu^{-} \to e^{-} \gamma$ & $(2,1)$ &
	$\text{Br}(\mu \to e\gamma) < 4.2 \times 10^{-13}$  \\
$\tau^{-} \to e^{-} \gamma$ & $(3,1)$ &
	$\text{Br}(\tau \to e\gamma) < 3.3 \times 10^{-8}$ \\
$\tau^{-} \to \mu^{-} \gamma$ & $(3,2)$ &
	$\text{Br}(\tau \to \mu\gamma) < 4.4 \times 10^{-8}$  \\ \hline
\end{tabular}
\caption{Summary of $\ell_i \to \ell_j \gamma$ process and the lower bound of experimental data~\cite{TheMEG:2016wtm, Adam:2013mnn}.}
\label{tab:Cif}
\end{table}

{\it $\ell_i\to\ell_j\gamma$ }:
Our relevant lepton flavor violation process ($\ell_i\to\ell_j\gamma$) comes from the same terms of anomalous magnetic moment at the one-loop level in principle. Each  of flavor dependent process has to satisfy the current upper bound, as can be seen in Table~\ref{tab:Cif}.
However the contribution from $y_S$ can be always negligible assuming the diagonal $y_S$. This is because this term does not contribute to the neutrino masses. Hence we consider the contribution from $y_L$ only.
Then the branching form is given as 
\begin{align}
\text{Br}(\ell_i\to\ell_j\gamma)&\approx
\frac{48\pi^3\alpha_{em}C_i}{G_F^2 m_{\ell_i}^2}\left(m_{\ell_i}^2 |a_R|^2+ m_{\ell_j}^2 |a_L|^2\right),\\
(a_R)_{ij}&=-
\sum_{a,b}^{1-3}\frac{f^*_{ia} f_{jb}}{16\pi^2}\int dxdydz\delta(x+y+z-1)xy\nn\\
&\left(\frac{\delta_{ab}}{\Delta[m_{\Sigma^{\pm\pm}_a},m_{\eta^\pm}]} 
+\frac{2\delta_{ab}}{\Delta[m_{\eta^\pm}, m_{\Sigma^{\pm\pm}_a}]} 
+\sum_{\kappa}^{1-6}\frac{(V_C^T)_{3+b,\kappa}(V_C)_{\kappa,3+a} }
{2 \Delta[m_{\eta^0},m_{\psi^\pm_\kappa}]}
\right),\\
(a_L)_{ij}&=-
\sum_{a,b'}^{1-3}\frac{f^*_{ia} f_{jb}}{16\pi^2}\int dxdydz\delta(x+y+z-1)xz\nn\\
&\left(\frac{\delta_{ab}}{\Delta[m_{\Sigma^{\pm\pm}_a},m_{\eta^\pm}]} 
+\frac{2\delta_{ab}}{\Delta[m_{\eta^\pm}, m_{\Sigma^{\pm\pm}_a}]}
+\sum_{\kappa}^{1-6}\frac{(V_C^T)_{3+b,\kappa}(V_C)_{\kappa,3+a} }
{2 \Delta[m_{\eta^0},m_{\psi^\pm_\kappa}]}\right),
\label{eq:g-2}
\end{align}
where 
\begin{align}
\Delta[m_1,m_2]&\equiv (x^2-x) m^2_{\ell_i} + x m_1^2+(y+z) m^2_2+xz(m^2_{\ell_i}-m^2_{\ell_j}), 
\end{align}
and ${ G_F}$ is Fermi constant, $\alpha_{em}$ is the fine structure constraint. $C_i\approx(1,1/5)$ for $i=(\mu,\tau)$.

\begin{table}[t]
\begin{tabular}{c|c|c|c|c} \hline
Nucleus $^A_Z N$ & $Z_{\rm eff}$ & $|F(-m^2_\mu)|$ & $|\Gamma_{\rm capt}(10^6{\rm sec}^{-1})$ & Experimental bounds (Future bound) \\ \hline
$^{27}_{13} Al$ & $11.5$ & $0.64$ & $0.7054$  & ($R_{Al}\lesssim10^{-16}$)~\cite{Hungerford:2009zz} \\
$^{48}_{22} Ti$ & $17.6$ & $0.54$ &$2.59$  & 
$R_{Ti}\lesssim 4.3\times 10^{-12}$~\cite{Dohmen:1993mp}\ 
($\lesssim 10^{-18}$ \cite{Hungerford:2009zz}) \\
$^{197}_{79} Au$ & $33.5$ & $0.16$ & $13.07$ & $R_{Au}\lesssim7\times 10^{-13}$ ~\cite{Bertl:2006up}  \\ 
$^{208}_{82} Pb$ & $34$ & $0.15$ & $13.45$   & $R_{Pb}\lesssim4.6\times 10^{-11}$~\cite{Honecker:1996zf}  \\ \hline
\end{tabular}
\caption{
Summary for the the $\mu\mathchar`-e$ conversion in various nuclei: 
$Z$, $Z_{\rm eff}$, $F(q)$, $\Gamma_{\rm capt}$, and the bounds on
the capture rate $R$.}
\label{tab:mue-conv}
\end{table}

{\it $\mu -e$ conversion}:
The $\mu-e$ conversion rate $R$ 
 is given by~\cite{Hisano:1995cp}
\begin{align}
R&=\frac{\Gamma(\mu\to e)}{\Gamma_{\rm capt}},\\
\Gamma(\mu\to e)&=
C_{\mu e}
\left[
\left|Z\left[(b_L)_{21} - (a_R)_{21} \right]
- (b_L)_{21} \frac{(2Z+N)A_u+(Z+2N)A_d}{2(s_{tw}c_{tw})^2}  \right|^2
+\left| Z (a_L)_{21} \right|^2
\right],
\end{align}
where $C_{\mu e}\equiv4\alpha_{\rm em}^5 \frac{Z^4_{\rm eff}|F(q)|^2 m^5_\mu}{Z}$, $A_u\equiv -\frac12-\frac43s_{tw}^2$, $A_d\equiv -\frac12+\frac23s_{tw}^2$, $\sin^2\theta_w\equiv s_{tw}^2\approx0.23$. The values of $\Gamma_{\rm capt}$, $Z$, $N$, $Z_{\rm eff}$, and $F(q)$ are summarized in Table~\ref{tab:mue-conv}.
$b^V_L$ in our model is given by
\begin{align}
&(b_L)_{21}=-\sum_{a,b=1}^3\frac{f_{1a} f^\dag_{b 2} }{(4\pi)^2}
\int dxdy dz\delta(x+y+z-1)(x+z)(1-x-z) \nn\\
&\times
\left[
\frac{2}{\Delta[m_{\eta^\pm},m_{\Sigma^{\pm\pm}}]} 
+\sum_{\kappa}^{1-6}\frac{(V_C^T)_{3+a,\kappa}(V_C^*)_{\kappa,3+b} }
{2 \Delta[m_{\eta^0},m_{\psi^\pm_\kappa}]}
\right].
\end{align}

{\it Muon anomalous magnetic moment} ($(g-2)_\mu$) is also given in terms of $a_{L/R}$ as
\begin{align}
{\Delta a_\mu\approx -{m_\mu^2}[(a_L)_{22}+(a_R)_{22}],}
\end{align}
which can be tested in the current experiments such as~\cite{Bennett:2006fi}.

\subsection{Dark matter}
Here we focus on the $\chi_R$ dominated DM candidate denoting $X \simeq \chi_R$, therefore all the cross sections proportional to $s_R^4$ are neglected due to $s_R<<1$. 
{The relevant interactions for calculating DM physics are obtained from scalar potential. In our analysis, they are parametrized as 
\begin{align}
-\mathcal{L} \supset & \sum_{A=1}^3 \mu_{2Xh_A} X X h_A + \sum_{A=1}^3 \mu_{h_A 2h_{SM}} h_A h_{SM} h_{SM} + \sum_{\alpha=1}^2 \mu_{XH_\alpha h_{SM}} X H_\alpha h_{SM} \nn \\
& + \lambda_{2X 2 h_{SM}} (X h_{SM})^2.
\end{align}
These interactions induce DM annihilation processes: $XX \to \bar f_{SM} f_{SM}$, $XX \to VV$ and $XX \to h_{SM} h_{SM}$ where $f_{SM}$ denotes SM fermions and $V = W, Z$.}
Then the relic density can be written by~\cite{Griest:1990kh}
\begin{align}
\Omega h^2\approx \frac{4.28\times 10^9 x_f^2}{ \sqrt{g^*} M_{pl}[(-3+4 x_f) a_{eff}+12 b_{eff}]},
\end{align}
where $g^*\approx100$, $M_P\approx 1.22\times 10^{19}$, $x_f\approx25$, and $a_{eff}$ and $b_{eff}$ are obtained by expanding the thermally averaged cross section as
\begin{align}
\sigma v_{rel}\approx \frac{1}{32\pi^2 s}\sum_{fin }\sqrt{1-\frac{4 m_{fin}^2}{s}}\int d\Omega |\bar M_{fin}|^2\approx a_{eff}+b_{eff} v^2_{rel},
\end{align}
where $\int d\Omega\equiv 2\pi \int_0^{\pi} d\theta \sin\theta$, and $ |\bar M_{fin}|^2\approx  |\bar M_{f_{SM}(\approx t,b,c,\tau)}|^2+   |\bar M_{V(=Z,W)}|^2+  |\bar M_{h_{SM}}|^2$, with
\begin{align}
& |\bar M_{f_{SM}}|^2\approx 8 C_f \left| \sum_{A=1}^3\frac{\mu_{2Xh_A} m_f (O_R)^T_{2A}}{v(s-m_{h_A}^2)}\right|^2 (s-4 m_f^2),\\
&  |\bar M_{V}|^2\approx C_V \left| \sum_{A=1}^3\frac{2\mu_{2Xh_A} G_V^{(A)}}{s-m_{h_A}^2}\right|^2 
\left(3 + \frac{s^2}{4m_V^4}-\frac{s}{m_V^2}\right)
, \quad (V=W, Z),\\
&  |\bar M_{h_{SM}}|^2\approx \frac12\left|
 \lambda_{2X2h_{SM}} +
\sum_{A=1}^3\frac{\mu_{2Xh_A} \mu_{h_A 2h_{SM}}} {s-m_{h_A}^2}+
\sum_{\alpha=1}^2
(\mu_{XH_\alpha h_{SM}})^2\left(\frac1{t-m_{H_\alpha}^2} + \frac1{u-m_{H_\alpha}^2} \right)
\right|^2.
\end{align}
Here $C_f=1$ for $f=\tau$, $C_f=3$ for $f=t,b,c$, $C_W=1$, $C_V=1/2$, $s,\ t,\ u$ are Mandelstam valuables, and 
\begin{align}
& G_W^{(A)}\equiv g_2^2 \left[ 2 v_\Delta (O_R)^T_{1A} + v  (O_R)^T_{2A}  \right],\quad
& G_Z^{(A)}\equiv \frac{2g_2^2}{c_{tw}^2} \left[ 2 v_\Delta (O_R)^T_{1A} + \frac v2  (O_R)^T_{2A}  \right].
\end{align}

{\it Direct detection}:
We have a spin independent scattering cross section with nucleon through $h_{1,2, 3}$ portal processes and its form is given by
\begin{align}
\sigma_N\approx 0.082\frac{m_N^4}{\pi v^2 M_X^2}
\left|
\sum_{A=1}^3
\frac{\mu_{2Xh_A (O_R)^T_{2A}}}{m_{h_A}^2}
\right|^2,
\end{align}
where the mass of nucleon, which is symbolized by $m_N$, is around 0.939 GeV. Recent LUX experiment in 2016 reported the lower bound on $\sigma_N\lesssim$2.2$\times$ 10$^{-46}$ cm$^2$ at 50 GeV mass range of DM at  the 90 \% confidential level~\cite{Akerib:2016vxi}.

\section{Numerical results}
Now that all the formulae have been provided, we have a numerical analysis.
Here we provide a benchmark point to satisfy all the constraints and discuss our predictions.
Once input parameters are given by 
\begin{align}
&(m_{h_1}, m_{h_{\rm SM}} ,m_{h_3}, m_a,m_C)\approx(488, 125, 418, 577, 442)\ {\rm [GeV]}, \quad M_{\Sigma} \approx{\rm diag.}(890, 795, 826)\ {\rm [GeV]} \nn\\
& M_{N_L}\approx{\rm diag.}(0.886, 0.417, 0.237)\ {\rm [GeV]},\quad
M_{N_R}\approx{\rm diag.}(0.409, 0.402, 0.247)\ {\rm [GeV]},\nn\\
& M_{N_\Sigma}\approx{\rm diag.}(14.7, 9.06, 7.36)\ {\rm [GeV]},\quad
M_{L}\approx{\rm diag.}(570, 509, 561)\ {\rm [GeV]},\nn\\
& (M^2)^{inert}_{\rm CP-even(odd)} \approx
\left[\begin{array}{cc}
2.18(3.36) \times10^5 & 4299 \\
4299 & 3.02(3.48)\times10^5 \\
\end{array}\right]\ {\rm [GeV^2]},
\nn\\
& (\alpha_1,\alpha_2,\alpha_3)\approx{\rm diag.}(3.61, 4.06, 2.38),\quad
(m_{\eta^0}(=m_{\eta^\pm}), m_{\chi^0},m_{\Delta^{\pm\pm}})\approx (970, 426, 959)\ {\rm [GeV]},\nn\\
& 
\delta M_{L} \approx{\rm diag}.(0.00786, 0.00368, 0.00215)\ {\rm [GeV]}, \quad (\delta m_{\chi^0\eta^0},\delta m_{\chi^0})\approx {\rm diag}.(0.905, 0.178)\ {\rm [GeV]}, \nn\\
& 
\mu_\eta \approx 420 \ {\rm GeV},\quad v_\Delta\approx 0.871 \ {\rm [GeV]},\quad \mu_{2X h_A} \approx (0.00237, 0.000175, 0.0642)\ {\rm [GeV]}, \nn\\
& \mu_{h_A 2 h_{SM}} \approx (0.741, 0.0608, 0.0311)\ {\rm [GeV]},\quad \mu_{XH_\alpha h_{SM}} \approx (3.66, 0.0404)\ {\rm [GeV]}, \nn\\
& ((\theta_R)_{23},  (\theta_R)_{13}, (\theta_R)_{12}, \theta_I,\theta_C)\approx (0.0676, 3.27, 0.0990, 0, 0),   \quad \lambda_{2X2h_{SM}}\approx 0.293, 
\end{align}
then the output physical values are found as 
\begin{align}
& M_X\approx 466\ {\rm GeV},\quad \Omega h^2\approx 0.120,\quad 
T^{0\nu\beta\beta}_{1/2}({\rm Ge})\approx 3.43 \times 10^{86}\ {\rm yr},\quad
T^{0\nu\beta\beta}_{1/2}({\rm Xe})\approx 1.89 \times 10^{86}\ {\rm yr},\nn\\
&T^{0\nu\beta\beta}_{1/2}({\rm Nd})\approx 5.21 \times 10^{84}\ {\rm yr},\quad
   T^{0\nu\beta\beta}_{1/2}({\rm Te})\approx 6.04 \times 10^{85}\ {\rm yr},\quad
   T^{0\nu\beta\beta}_{1/2}({\rm Se})\approx 9.73 \times 10^{85}\ {\rm yr},\nn\\
 &
 BR(\mu\to e\gamma)\approx 2.52 \times 10^{-20},\quad  BR(\tau\to e\gamma)\approx 1.59\times 10^{-22},\quad
   BR(\tau\to \mu\gamma)\approx 6.60 \times 10^{-23},\nn\\
 &
 R_  {\rm Al}\approx 2.12 \times 10^{-22},\quad  R_  {\rm Ti}\approx 3.88 \times 10^{-22},\quad
R_  {\rm Au}\approx 3.34\times 10^{-22}, \quad R_  {\rm Pb}\approx 3.17 \times 10^{-22},\nn\\
&
\sigma_N\approx 1.37 \times 10^{-53}\ {\rm cm^2},\quad \Delta S\approx 0.0415 \quad \Delta T\approx0.00689,\quad
\Delta a_\mu\approx 2.00 \times 10^{-16},
\end{align}
where Yukawa coupling $f$ is determined by the Casas-Ibarra parametrization, and the typical order is ${\cal O}(10^{-4})$. 

Here several remarks are in order:
\begin{enumerate}

\item
{\it Relic density of DM} can be controlled by the output quartic coupling $\lambda_{2X2h_{SM}}$ without affecting the other phenomenologies. Thus any value of the DM mass is possible, depending on its coupling. Also trilinear terms $\mu_{2X h_A},\ \mu_{h_A 2 h_{SM}},\ \mu_{XH_\alpha h_{SM}}$ can also be valid to control the relic density, although their too much heavier terms  are conflict with the direct detection search.

\item
Although there are {\it non standard interactions of neutrinoless double beta decay at the box type one-loop level} in our model,
the constraints are much  weaker than the current bounds. Thus its dominant contribution arises from the standard interaction.
{The main reason of this smallness comes from the tiny Yukawa coupling $f={\cal O}(10^{-4})$.}~\footnote{If one selects a specific parametrization, one might find the enhanced value of $\Delta a_\mu$ without conflict with LFVs. However since this is something like a fine-tuning, we do not consider such a specific case.} Thus all the other phenomenologies related to this coupling such as $\ell_i\to \ell_j\gamma$, $\mu-e$ conversion, and muon $g-2$, are also very small and easily be evaded  the current experimental bounds.

\item
It might be worthwhile mentioning the future testability of $\mu-e$ conversion. Especially the sensitivity of $R$ with nucleons $Al$ and $Ti$ will be significantly improved by $R_{Al}\lesssim10^{-16}$ and $R_{Ti}\lesssim10^{-18}$. Although our results are still weaker than these future bounds,
{\it this could be one of the possibilities to discriminate the other radiative neutrino models, because their typical lowest bounds are at most ${\cal O}(10^{-15})-{\cal O}(10^{-16})$.}

\item
Due to the constraints of oblique parameters, masses related to this constraints cannot be taken to be so freely.
However one might control their masses by changing their mixings. In this case, one has to take care of the mixings between the SM Higgs and the other two heavier CP-even bosons, which are restricted by the 8 TeV collider at Large Hadron Collider (LHC), and also notice that {mass insertion approximations are not valid. Thus more complicated analyses has to be achieved.}

\end{enumerate}


\section{Conclusions}
We have studied a one-loop induced radiative neutrino model, in which we have shown an allowed bench mark point to satisfy
the observed neutrino masses, LFVs, and the relic density of DM satisfying the current upper bound  on the spin independent scattering with nucleon. Also we have shown the non-standard contribution to the muon $g-2$ and neutrinoless double beta decay in our model.
However, due to the small Yukawa coupling whose typical order is $10^{-4}$, all the constraints such as LFVs are much weaker than the current upper bounds. Thus muon $g-2$ and neutrinoless double beta decay are also very small in the typical range.
But we might discriminate the other radiative neutrino models from our model by the future experiment of $\mu-e$ conversion.

{Before closing, we would like to comment possible collider signal of the model. Since we have scalar and exotic leptons in $SU(2)_L$ triplet with hypercharge $\pm1$, there are doubly charged Higgs and leptons which have mass of $\lesssim 1$ TeV and would provide specific signature. They can be produced via electroweak process at the LHC. Then doubly charged Higgs dominantly decays into same sign $W$ boson pair or singly charged exotic leptons decaying into DM and SM fermions, depending on Yukawa couplings and triplet VEV. On the other hand, doubly charged leptons also decay into SM charged leptons and inert Higgs decaying into DM and SM fermions. Thus these doubly charged particles have cascade decay mode providing DM plus SM fermions. Since detailed analysis is beyond the scope of this paper it will be given in elsewhere.}

\section*{Acknowledgments}
\vspace{0.5cm}
H.O. expresses his sincere gratitude toward all the KIAS members, Korean cordial persons, foods, culture, weather, and all the other things.
This work was supported by the Korea Neutrino Research Center which is established by the National Research Foundation of Korea(NRF) grant funded by the Korea government(MSIP) (No. 2009-0083526) (Y.O.).

\begin{appendix}

\section{New particle contribution to vacuum polarization diagram}
Here we summarize contributions to $\Pi_{\pm}(q^2)$, $\Pi_{33}(q^2)$, $\Pi_{3Q}(q^2)$ and  $\Pi_{QQ}(q^2)$ in Eq.~(\ref{eq:piZ}) and (\ref{eq:piW}) from new particles in our model.

\noindent
{\bf For Vacuum polarization diagram of $W^\pm W^\pm$} \\
\noindent
{\it $\psi^\pm_\kappa$-$\psi_n$ loop contribution} :
\begin{align}
\Pi_\pm^{\psi^\pm_\kappa \psi_n}(q^2) = - \frac{1}{2 (4 \pi)^2} \left[ (\omega_{n \kappa} \omega_{n \kappa} + \omega'_{n \kappa } \omega'_{n \kappa} )F_\omega (q^2, m_{\psi_n}^2, m^2_{\psi^\pm_\kappa}) 
+ (\omega_{n \kappa} \omega'_{n \kappa} + \omega'_{n \kappa} \omega_{n \kappa} ) \bar F_\omega (q^2, m_{\psi_n}^2, m^2_{\psi^\pm_\kappa})   \right]. 
\end{align}
where 
\begin{align}
&\omega_{n \kappa} = \sum_{a=1}^3 \left[ \frac{1}{\sqrt{2}} (V_N)_{n a} (V_C^T)_{a \kappa} + (V_N)_{n, 6+  a}(V_C^T)_{3+a, \kappa} \right] \nonumber \\
&\omega'_{n \kappa}  = \sum_{a=1}^3 \left[ \frac{1}{\sqrt{2}} (V_N)_{n, 3+ a} (V_C^T)_{a \kappa} + (V_N)_{n, 9+ a}(V_C^T)_{3+a, \kappa} \right] \nonumber \\
&F_\omega (q^2, m_{P}^2, m^2_{Q}) = \int dx dy \delta (1-x-y) (\Upsilon - \ln \Delta_{PQ}) [2x(1-x) q^2 - x m_{P}^2 - y m_{Q}^2] \nonumber \\
& \bar F_\omega (q^2, m_{P}^2, m^2_{Q}) = \int dx dy \delta (1-x-y) (\Upsilon - \ln \Delta_{PQ}) m_{P} m_{Q} \nonumber \\
& \Delta_{PQ} = -q^2 x(1-x) + x m_{P}^2 + y m_{Q}^2, \quad \Upsilon = \frac{2}{\epsilon} - \gamma - \ln (4 \pi).
\end{align}

\noindent
{\it $\psi^\pm_\kappa$-$\Sigma^{\pm \pm}_a$ loop contribution} :
\begin{align}
& \Pi_\pm^{\psi^\pm_\kappa \Sigma^{\pm \pm}_a}(q^2) = - \frac{1}{(4 \pi)^2}  (V_C)_{\kappa,3+a} (V^T_C)_{3+a,\kappa} F'_\omega(q^2, m_{\psi^\pm_\kappa}^2, m_{\Sigma_a}^2), \nonumber \\
&  F'_\omega(q^2, m_{P}^2, m^2_{Q}) = F_\omega (q^2, m_{P}^2, m^2_{Q}) + \bar F_\omega (q^2, m_{P}^2, m^2_{Q}).
\end{align} 

\noindent
{\it $H_\alpha^\pm$-$h_A$ loop contribution} :
\begin{equation}
\Pi_\pm^{H_\alpha^\pm h_A}(q^2) = \frac{2}{(4 \pi)^2} \left[ \frac{1}{\sqrt{2}} (O^T_C)_{1 \alpha} (O^T_R)_{1A} + \frac{1}{2} (O_C^T)_{2 \alpha} (O_R^T)_{2 A} \right]^2 G(q^2, m_{h_A}^2, m_{H_\alpha^\pm}^2).
\end{equation}
where 
\begin{align}
 G(q^2, m_P^2, m_Q^2) =  \int dx dy \delta (1-x-y) \Delta_{PQ} [\Upsilon+1 - \ln \Delta_{PQ}].
\end{align}

\noindent
{\it $H_\alpha^\pm$-$a_\beta$ loop contribution} :
\begin{align}
& \Pi_\pm^{H_\alpha^\pm a_\beta}(q^2) = \frac{2}{(4 \pi)^2} \left[ \frac{1}{\sqrt{2}} (O^T_C)_{1 \alpha} (O^T_I)_{1\beta} + \frac{1}{2} (O_C^T)_{2 \alpha} (O_I^T)_{2 \beta} \right]^2 G(q^2, m_{a_\beta}^2, m_{H_\alpha^\pm}^2).
\end{align}

\noindent
{\it $\eta^\pm$-$A_\alpha$ loop contribution} :
\begin{align}
\Pi_\pm^{\eta^\pm A_\alpha}(q^2) = \frac{1}{2(4 \pi)^2} (V_I^T)_{1 \alpha} (V_I^T)_{1 \alpha} G(q^2, m_{\eta^\pm}^2, m_{A_\alpha}^2). 
\end{align}

\noindent
{\it $\Delta^{\pm \pm}$-$H^\pm_\alpha$ loop contribution} :
\begin{align}
\Pi_\pm^{\Delta^{\pm\pm} H^\pm_\alpha}(q^2) = \frac{2}{(4 \pi)^2} (O_C^T)_{1\alpha} (O_C^T)_{1\alpha} G(q^2, m_{H^\pm_\alpha}^2, m_{\Delta}^2). 
\end{align}

\noindent
{\it $\eta^\pm$-$H_\alpha$ loop contribution} :
\begin{align}
\Pi_\pm^{\eta^\pm H_\alpha}(q^2) = \frac{1}{2(4 \pi)^2} (V_R^T)_{1 \alpha} (V_R^T)_{1 \alpha} G(q^2, m_{\eta^\pm}^2, m_{H_\alpha}^2). 
\end{align}

\noindent
{\it $h_A$ loop contribution} :
\begin{equation}
\Pi^{h_A}_\pm (q^2) = - \frac{1}{(4\pi)^2} \left[ \frac{1}{2} (O_R^T)_{1A} (O^T_R)_{1A} + \frac{1}{4} (O^T_R)_{2A} (O_R^T)_{2A} \right]^2 H (m_{h_A}^2),
\end{equation}
where
\begin{equation}
H(m_P^2) = m_P^2 [\Upsilon + 1 - \ln m_P^2].
\end{equation}

\noindent
{\it $a_\alpha$ loop contribution} :
\begin{equation}
\Pi^{a_\alpha}_\pm (q^2) = - \frac{1}{(4\pi)^2} \left[ \frac{1}{2} (O_I^T)_{1 \alpha} (O^T_I)_{1 \alpha} + \frac{1}{4} (O^T_I)_{2 \alpha} (O_I^T)_{2 \alpha} \right]^2 H (m_{a_\alpha}^2).
\end{equation}

\noindent
{\it $H_\alpha$ loop contribution} :
\begin{align}
 \Pi^{H_\alpha}_\pm (q^2) = - \frac{1}{4(4\pi)^2} (V_R^T)_{1 \alpha} (V_R^T)_{1 \alpha} H (m_{H_\alpha}^2).
\end{align}

\noindent
{\it $A_\alpha$ loop contribution} :
\begin{align}
 \Pi^{A_\alpha}_\pm (q^2) = - \frac{1}{4(4\pi)^2} (V_I^T)_{1 \alpha} (V_I^T)_{1 \alpha} H (m_{A_\alpha}^2).
\end{align}

\noindent
{\it $H^\pm_\alpha$ loop contribution} :
\begin{equation}
 \Pi^{H^\pm_\alpha}_\pm (q^2) = - \frac{1}{(4\pi)^2} \left[ 2 (O_C^T)_{1\alpha} (O^T_C)_{1 \alpha} + \frac{1}{2} (O^T_C)_{2 \alpha} (O_C^T)_{2 \alpha}\right]^2 H (m_{H^\pm_\alpha}^2). 
\end{equation}

\noindent
{\it $\eta^\pm$ loop contribution} :
\begin{align}
 \Pi^{\eta^\pm}_\pm (q^2) = - \frac{1}{2(4\pi)^2} H (m_{\eta^\pm}^2).
\end{align}

\noindent
{\it $\Delta^{\pm \pm}$ loop contribution} :
\begin{align}
 \Pi^{\Delta^{\pm \pm}}_\pm (q^2) = - \frac{1}{(4\pi)^2} H (m_{\Delta}^2).
\end{align}


\noindent
{\bf For Vacuum polarization diagram of $ZZ$} \\
\noindent
{\it $\psi_n$-$\psi_m$ loop contribution} :
\begin{align}
& \Pi_{33}^{\psi_n \psi_m}(q^2) = - \frac{1}{2 (4 \pi)^2} X_{nm} X_{mn}  F_\omega (q^2, m_{\psi_n}^2, m^2_{\psi_m}), \nonumber \\
& X_{nm} = \sum_{a=1}^{3} \left[ \frac{1}{2} (V_N)_{n a} (V_N^T)_{a m} + \frac{1}{2} (V_N)_{n, 3+ a}(V_N^T)_{3+a, m} 
+(V_N)_{n, 6+ a}(V_N^T)_{6+a, m} + (V_N)_{n, 9+ a}(V_N^T)_{9+a, m} \right].  
\end{align}

\noindent
{\it $\psi^\pm_\kappa$-$\psi^\pm_\zeta$ loop contribution} :
\begin{align}
& \Pi_{33}^{\psi^\pm_\kappa \psi^\pm_\zeta}(q^2) = - \frac{1}{4 (4 \pi)^2} C_{\kappa \zeta} C_{\zeta \kappa}  F'_\omega(q^2, m_{\psi^\pm_\kappa}^2, m^2_{\psi^\pm_\zeta}), \nonumber \\
& \Pi_{3Q}^{\psi^\pm_\kappa \psi^\pm_\zeta}(q^2) = - \frac{1}{ (4 \pi)^2} \left[\frac{1}{2} C_{\kappa \zeta} C_{\zeta \kappa} +\frac{1}{4}(C_{\kappa \zeta}D_{\zeta \kappa}+D_{\kappa \zeta}C_{\zeta \kappa}) \right] F'_\omega(q^2, m_{\psi^\pm_\kappa}^2, m^2_{\psi^\pm_\zeta}), \nonumber \\
& \Pi_{QQ}^{\psi^\pm_\kappa \psi^\pm_\zeta}(q^2) = - \frac{1}{ (4 \pi)^2} \left[ C_{\zeta \kappa} C_{\kappa \zeta} +C_{\kappa \zeta}D_{\zeta \kappa}+D_{\kappa \zeta}C_{\zeta \kappa} + D_{\kappa \zeta} D_{\zeta \kappa} \right] F'_\omega(q^2, m_{\psi^\pm_\kappa}^2, m^2_{\psi^\pm_\zeta}).
\end{align}
where 
\begin{align}
C_{\kappa \zeta} = \sum_{a=1}^3 (V_C)_{\kappa a} (V_C^T)_{a \zeta}, \quad D_{\kappa \zeta}  = \sum_{a=1}^3 (V_C)_{\kappa, 3+ a} (V_C^T)_{3+a, \zeta}.
\end{align}

\noindent
{\it $\Sigma^{\pm\pm}_a$-$\Sigma^{\pm\pm}_a$ loop contribution} :
\begin{equation}
 \Pi_{33}^{\Sigma^{\pm \pm}_a}(q^2) = \frac{1}{2} \Pi_{3Q}^{\Sigma^{\pm \pm}_a}(q^2) = \frac{1}{4} \Pi_{QQ}^{\Sigma^{\pm \pm}_a}(q^2) = - \frac{1}{(4 \pi)^2} F'_\omega(q^2, m_{\Sigma_a}^2, m^2_{\Sigma_a}). \nonumber \\
\end{equation}

\noindent
{\it $h_A$-$a_\alpha$ loop contribution} :
\begin{equation}
\Pi_{33}^{h_A a_\alpha}(q^2) = \frac{2}{(4\pi)^2} \left[ (O_R^T)_{1A} (O_I^T)_{1\alpha} + \frac{1}{2} (O^T_R)_{2A} (O_I^T)_{2\alpha} \right]^2 G(q^2, m_{h_A}^2, m_{a_\alpha}^2), 
\end{equation}

\noindent
{\it $H_\alpha^\pm$-$H_\beta^\pm$ loop contribution} :
\begin{align}
\Pi_{33}^{H_\alpha^\pm H_\beta^\pm} (q^2) &= \frac{1}{2(4\pi)^2} B_{\alpha \beta} B_{\beta \alpha} G(q^2, m_{H_\alpha^\pm}^2, m_{H_{\beta}^\pm}^2), \nonumber \\
\Pi_{3Q}^{H_\alpha^\pm H_\beta^\pm} (q^2) &= \frac{1}{2(4\pi)^2} \left[ - \frac{1}{2} \left( A_{\alpha \beta} B_{\beta \alpha} + B_{\alpha \beta} A_{\beta \alpha} \right) + 2 B_{\alpha \beta} B_{\beta \alpha} \right] G(q^2, m_{H_\alpha^\pm}^2, m_{H_{\beta}^\pm}^2), \nonumber \\
\Pi_{QQ}^{H_\alpha^\pm H_\beta^\pm} (q^2) &= \frac{1}{2(4\pi)^2} \left[A_{\alpha \beta} A_{\beta \alpha} - 2 \left( A_{\alpha \beta} B_{\beta \alpha} + B_{\alpha \beta} A_{\beta \alpha} \right) + 4 B_{\alpha \beta} B_{\beta \alpha} \right] G(q^2, m_{H_\alpha^\pm}^2, m_{H_{\beta}^\pm}^2),
\end{align}
where 
\begin{align}
A_{\alpha \beta} = (O^T_C)_{1\alpha} (O_C^T)_{1\beta}, \quad B_{\alpha \beta} = (O_C^T)_{2 \alpha} (O_C^T)_{2 \beta}.
\end{align}

\noindent
{\it $\Delta^{\pm \pm}$-$\Delta^{\pm \pm}$ loop contribution} :
\begin{equation}
\Pi_{33}^{\Delta^{\pm\pm} \Delta^{\pm\pm}} (q^2) = \frac{1}{2} \Pi_{3Q}^{\Delta^{\pm\pm} \Delta^{\pm\pm}} (q^2) 
= \frac{1}{4} \Pi_{QQ}^{\Delta^{\pm\pm} \Delta^{\pm\pm}} (q^2) = \frac{2}{(4\pi)^2} G(q^2, m_\Delta^2, m_\Delta^2). 
\end{equation}

\noindent
{\it $\eta^{\pm}$-$\eta^{\pm}$ loop contribution} :
\begin{equation}
\Pi_{33}^{\eta^{\pm} \eta^{\pm}} (q^2) = \frac{1}{2} \Pi_{3Q}^{\eta^{\pm} \eta^{\pm}} (q^2) = \frac{1}{4} \Pi_{QQ}^{\eta^{\pm} \eta^{\pm}} (q^2)= \frac{1}{2(4\pi)^2} G(q^2, m_{\eta^\pm}^2, m_{\eta^\pm}^2).
\end{equation}

\noindent
{\it $H_\alpha$-$A_\beta$ loop contribution} :
\begin{equation}
\Pi_{33}^{H_\alpha A_\beta} (q^2) =  \frac{1}{2(4\pi)^2} \left[(V_R^T)_{1 \alpha} (V_I^T)_{1 \beta} \right]^2 G(q^2, m_{H_\alpha}^2, m_{A_\beta}^2).
\end{equation}

\noindent
{\it $h_A$ loop contribution} :
\begin{equation}
\Pi_{33}^{h_A} (q^2) = - \frac{1}{(4 \pi)^2} \left[ (O_R^T)_{1A} (O_R^T)_{1A} + \frac{1}{4} (O_R^T)_{2A} (O_R^T)_{2A} \right] H(m_{h_A}^2).
\end{equation}

\noindent
{\it $a_\alpha$ loop contribution} :
\begin{equation}
\Pi_{33}^{a_\alpha}(q^2)  = - \frac{1}{(4 \pi)^2} \left[ (O_I^T)_{1 \alpha} (O_I^T)_{1 \alpha} + \frac{1}{4} (O_I^T)_{2 \alpha} (O_I^T)_{2 \alpha} \right] H(m_{a_\alpha}^2).
\end{equation}

\noindent
{\it $H^\pm_\alpha$ loop contribution} :
\begin{align}
\Pi_{33}^{H_\alpha^\pm} (q^2) &= - \frac{1}{2 (4 \pi)^2} B_{\alpha \alpha} H (m_{H^\pm_\alpha}^2), \nonumber \\
\Pi_{3Q}^{H_\alpha^\pm} (q^2) &= - \frac{1}{ (4 \pi)^2} B_{\alpha \alpha} H (m_{H^\pm_\alpha}^2), \nonumber \\
\Pi_{QQ}^{H_\alpha^\pm} (q^2) &= - \frac{2}{ (4 \pi)^2}[ A_{\alpha \alpha} + B_{\alpha \alpha}]H (m_{H^\pm_\alpha}^2), 
\end{align}

\noindent
{\it $\Delta^{\pm \pm}$ loop contribution} :
\begin{equation}
\Pi_{33}^{\Delta^{\pm \pm}}(q^2) = \frac{1}{2} \Pi_{3Q}^{\Delta^{\pm \pm}}(q^2) = \frac{1}{4} \Pi_{QQ}^{\Delta^{\pm \pm}}(q^2) = - \frac{2}{(4\pi)^2} H(m_{\Delta}^2).
\end{equation}

\noindent
{\it $\eta^{\pm}$ loop contribution} :
\begin{equation}
\Pi_{33}^{\eta^{\pm}}(q^2) = \frac{1}{2} \Pi_{3Q}^{\eta^{ \pm}}(q^2) = \frac{1}{4} \Pi_{QQ}^{\eta^{ \pm}}(q^2) = - \frac{1}{2(4\pi)^2} H(m_{\eta^\pm}^2).
\end{equation}

\noindent
{\it $H_\alpha$ loop contribution} :
\begin{equation}
\Pi_{33}^{H_\alpha}(q^2) = - \frac{1}{4(4\pi)^2} (V_R^T)_{1\alpha} (V_R^T)_{1\alpha} H(m^2_{H_\alpha}).
\end{equation}

\noindent
{\it $A_\alpha$ loop contribution} :
\begin{equation}
\Pi_{33}^{A_\alpha}(q^2) = - \frac{1}{4(4\pi)^2} (V_I^T)_{1\alpha} (V_I^T)_{1\alpha} H(m^2_{A_\alpha}).
\end{equation}


\section{Loop functions}

We define the functions $F_i$'s and $F_{3,2}$  as follows. 
 
For $n \ge 1$: 
\begin{align}
& F_2(\alpha, n, A, B) \equiv \int_0^1 dx \frac{x^\alpha}{(A x +B)^n}, 
\\
& F_3(\alpha, n, A, B, C) \equiv \int_0^1 dx\int_0^{1-x} dy \frac{x  y^\alpha}{(A x +B y +C)^n},  
\\
& F_4(\alpha, n, A, B,C, D) \equiv \int_0^1 dx\int_0^{1-x} dy\int_0^{1-x-y} dz \frac{x y z^\alpha}{(A x +B y+ C z +D)^n}, 
\\
& F_{3, 2}(\alpha, \beta, n, A, B, C) \equiv \int_0^1 dx\int_0^{1-x} dy \frac{x^\alpha  y^\beta}{(A x +B y +C)^n}. 
\end{align}

For $n=0$; 
\begin{align}
 F_2(\alpha, 0, A, B) \equiv \int_0^1 dx x^\alpha \ln (A x +B).  
\end{align}

For $n=-1$; 
\begin{align}
 F_2(0, -1, A, B) \equiv \int_0^1 dx x \ln (A x +B). 
\end{align}
Where n's and $\alpha$'s are integers. 

$F_2$ are calculated as follows. 

For $n\ge 2$
\begin{align}
& F_2(\alpha, n, A, B) = \frac1{A^\alpha} \left(F_2(0,n-\alpha, A,B) 
 - \sum^{\alpha-1}_{\beta=0} 
 \left(
 \begin{array}{c}
 \alpha \\
 \beta 
 \end{array}
 \right)
A^\beta B^{\alpha-\beta} F_2(\beta,n,A,B)\right),
\end{align}
Where $\alpha \ge 1$. 
In the cases of $n=-1, 0, 1$, $F_2$'s are as follows. 
\begin{align}
& F_2(\alpha, 0, A, B) = \frac{-1}{A(n-1)} \left( \frac1{(A+B)^{n-1}} - \frac1{B^{n-1}}\right), \\
& F_2(\alpha, 1, A, B) = \frac1A \left( F_2(0, n-1, A, B) - B F_2(0, n, A, B) \right), \\
& F_2(0, 1, A, B) = \frac1A  \ln\left(\frac{A +B}{B } \right), \\
& F_2(0, 0, A, B) = \frac{A+B}{A}  \ln\left(A +B \right) - \frac{B}{A}  \ln\left(B \right)-1, \\
& F_2(0, -1, A, B) = \frac{1}{2 A^2}\left( (A+B)^2 \ln\left(A +B \right) - B^2 \ln\left(B \right)- \frac{(A+B)^2}{2} + \frac{B^2}{2} \right). 
\end{align}
%

$F_3$, $F_{3,2}$ and $F_4$ can be written by 
$F_2$'s functions.

\begin{align}
& F_3(0,n,A,B,C)=-\frac{1}{B(n-1)}\left(F_2(1,n-1,A-B,B+C) -F_2(1,n-1,A,C)\right) \hspace{0.5cm} (n \ge 2). 
\nonumber\\ \\
& F_3(0,1,A,B,C)=-\frac{1}{B}\left(F_2(0,-1,A-B,B+C) -F_2(0,-1,A,C)\right).
\end{align}
\begin{align}
& F_3(1,n,A,B,C)=\frac{1}{B^2}\left(-\frac{1}{n-2} F_2(1,n-2,A-B,B+C) 
 +\frac{1}{n-1}\left(A F_2(2,n-1,A-B,B+C) 
\right.\right.\nonumber\\ & \left.\left.
 +C F_2(1,n-1,A-B,B+C)\right)
 +\frac{1}{(n-2)(n-1)} F_2(1,n-2,A,C)\right) \hspace{0.5cm} (n \ge 3). 
\\
& F_3(1,2,A,B,C)=\frac{1}{B^2}\left(F_2(0,-1,A-B,B+C)-F_2(0,-1,A,C)+A F_2(2,1,A-B,B+C)
 \right.\nonumber\\ & \left.
  +C F_2(1,1,A-B,B+C) -A F_2(2,1,A,C) -C F_2(1,1,A,C)
\right). \\
& F_3(1,1,A,B,C)=\frac{1}{6 B}-\frac{A}{B^2}\left(F_2(2,0,A-B,B+C) -F_2(2,0,A,C)\right)
\nonumber\\ &
 -\frac{C}{B^2}\left(F_2(1,0,A-B,B+C) -F_2(1,2,A,C)\right)
\end{align}
\begin{eqnarray}
F_3(2,n,A,B,C)&=&\frac{1}{B^3}\left(\frac{-1}{n-3}F_2(1,n-3,A-B,B+C)
 +\frac{2}{n-2}\left(A F_2(2,n-2,A-B,B+C)
  \right.\right. \nonumber\\ && \left.\left.
  +C F_2(1,n-2,A-B,B-C) \right)
-\frac{1}{n-1}\left(A^2 F_2(3,n-1,A-B,B+C)
  \right.\right. \nonumber\\ && \left.\left.
  +2 A C F_2(2,n-1,A-B,B+C)
  +C^2 F_2(1,n-1,A-B,B+C)\right)
  \right.\nonumber\\ && \left.
  \frac{2}{(n-3)(n-2)(n-1)} F_2(1,n-3,A,C)
\right) \hspace{0.5cm} (n \ge 4). 
\\
F_3(2,3,A,B,C)&=&\frac{1}{B^3}\left(F_2(1,0,A-B,B+C)-F2(1,0,A,C)\right)
\nonumber\\ &&
 +\frac{2}{B^3}\left(A \left(F_2(2,1,A-B,B+C) -F_2(2,1,A,C)\right)
 \right.\nonumber\\ && \left.
 +C \left(F_2(1,1,A-B,B+C) -F_2(1,1,A,C)\right) \right)
 \nonumber\\ &&
 -\frac{1}{2 B^2}\left(A^2\left(F_2(3,2,A-B,B+C)-F_2(3,2,A,C)\right)
 \right.\nonumber\\ && \left.
 +2 A C \left(F_2(2,2,A-B,B+C) -F_2(2,2,A,C)\right) 
 \right.\nonumber\\ && \left.
 +C^2\left(F_2(1,2,A-B,B+C)-F_2(1,2,A,C)\right)
  \right), 
\\
F_3(2,2,A,B,C)&=&\frac{1}{6 B^2} -\frac{1}{B^3}\left(
 2 A \left(F_2(2,0,A-B,B+C)-F_2(2,0,A,C)\right)
  \right.\nonumber\\ && \left.
 +2 C\left(F_2(1,0,A-B,B+C)-F_2(1,0,A,C)\right)
  \right.\nonumber\\ && \left.
 -A\left(F_2(3,1,A-B,B+C)-F_2(3,1,A,C)\right)
  \right.\nonumber\\ && \left.
 -2 A C\left(F_2(2,1,A-B,B+C)-F_2(2,1,A,C)\right)
  \right.\nonumber\\ && \left.
 -C^2\left(F_2(1,1,A-B,B+C)-F_2(1,1,A,C)\right)
\right), 
\\
F_3(2,1,A,B,C)&=&\frac{1}{6A}-\frac{1}{A^2}\left(B\left(F_2(3,0,A-B,B+C)-F_2(3,0,A,C)\right)
  \right.\nonumber\\ && \left.
+C\left(F_2(2,0,A-B,B+C)-F_2(2,0,A,C)\right)
\right). 
\end{eqnarray}

\begin{eqnarray}
F_3(3,n,A,B,C)=&\frac{1}{A}\left(\frac{-1}{A(n-2)}\left(F_2(3,n-2,B-A,A+C)-F_2(3,n-2,B,C)\right)
  \right.\nonumber\\ & \left.
  +\frac{1}{A(n-1)}\left(B\left(F_2(4,n-1,B-A,A+C)-F_2(4,n-1,B,C)\right)
  \right.\right.\nonumber\\ & \left.\left.
  +C \left(F_2(3,n-1,B-A,A+C)-F_2(3,n-1,B,C)\right)\right)
  \right) \hspace{0.5cm}(n\ge 3), \nonumber \\
\end{eqnarray}

\begin{eqnarray}
F_3(\alpha,2,A,B,C)&=&\frac{1}{A^2}\left(F_2(\alpha,0,B-A,A+C)-F_2(\alpha,0,B,C)
  \right.\nonumber\\ && \left.
  -B\left(F_2(\alpha+1,1,B-A,A+C)-F_2(\alpha+1,1,B,C)\right)
  \right.\nonumber\\ && \left.
 -C\left(F_2(\alpha,1,B-A,A+C)-F_2(\alpha,1,B,C)\right)
\right), 
\\
F_3(\alpha,1,A,B,C)&=&\frac{1}{A (\alpha+1)(\alpha+2)}+\frac{1}{A}\left(
  -B\left(F_2(\alpha+1,0,B-A,A+C)-F_2(\alpha+1,0,B,C)\right)
  \right.\nonumber\\ && \left.
 -C\left(F_2(\alpha,0,B-A,A+C)-F_2(\alpha,0,B,C)\right)
\right). 
\end{eqnarray}

\begin{align}
& F_4(1,n,A,B,C,D) \nn \\
&=\frac{1}{C^2}\left(
\frac{-1}{n-2}F_3(1,n-2,A-C,B-C,C+D) +\frac{1}{n-1}\left(D F_3(1,n-1,A-C,B-C,C+D)
  \right.\right.\nonumber\\ & \left.\left.
 + A F_3(2,n-1,B-C,A-C,C+D)  +B F_3(2,n-1,A-C,B-C,C+D)\right)
   \right.\nonumber\\ & \left.
  +\frac{1}{(n-2)(n-1)}F_3(1,n-2,A,B,D)
\right), 
\end{align}
\begin{align}
& F_4(2,n,A,B,C,D)= \nn \\
&\frac{1}{C^3(3-n)}\left(F_3(1,n-3,A-C,B-C,C+D) -F_3(1,n-3,A,B,D)\right)
 \nonumber\\ &
 -\frac{1}{C^2}\left(\frac{2 A}{C (2-n)}\left(F_3(2,n-2,B-C,A-C,C+D) -F_3(2,n-2,A,B,D)\right)
 \right. \nonumber\\ & \left.
 +\frac{2 B}{C(2-n)}\left(F_3(2,n-2,A-C,B-C,C+D) -F_3(2,n-2,A,B,D)\right)
 \right. \nonumber\\ & \left.
 +\frac{2 D}{C(2-n)}\left(F_3(1,n-2,A-C,B-C,C+D) -F_3(1,n-2,A,B,D)\right)
 \right. \nonumber\\ & \left.
 -\frac{A^2}{C(1-n)}\left(F_3(3,n-1,B-C,A-C,C+D) -F_3(3,n-1,B,A,D)\right)
 \right. \nonumber\\ & \left.
 -\frac{B^2}{C(1-n)}\left(F_3(3,n-1,A-C,B-C,C+D) -F_3(3,n-1,A,B,D)\right)
 \right. \nonumber\\ & \left.
 -\frac{D^2}{C(1-n)}\left(F_3(3,n-1,A-C,B-C,C+D) -F_3(3,n-1,A,B,D)\right)
 \right. \nonumber\\ & \left.
 -\frac{2 A D}{C(1-n)}\left(F_3(2,n-1,B-C,A-C,C+D) -F_3(2,n-1,B,A,D)\right)
 \right. \nonumber\\ & \left.
 -\frac{2 B D}{C(1-n)}\left(F_3(2,n-1,A-C,B-C,C+D) -F_3(2,n-1,A,B,D)\right)
  \right. \nonumber\\ & \left.
 -\frac{2 A B}{C(1-n)}\left(F_{3,2}(2,2,n-1,A-C,B-C,C+D)-F_{3,2}(2,2,n-1,A,B,D)\right)
 \right)
 \nonumber\\ &(n\ge2). 
\end{align}

%
\begin{align}
F_{3,2}(2, 2, n, A, B, C)=&\frac{1}{B^3 (3-n)}\left(F_2(2, n-3, A - B, B+C) - F_2(2, n-3, A, C)\right)
\nonumber\\
 &-\frac{2}{B^3(2-n)}\left(A \left(F_2(3, n-2, A-B, B+C) - F_2(3, n-2, A, C)\right)
\right.\nonumber\\ & \left.
  +C\left( F_2(2, n-2, A-B, B+C) - F_2(2, n-2, A, C) \right)\right)
\nonumber\\
 &+\frac{1}{B^3 (1-n)}\left( A^2 \left(F_2(4, n-1, A-B, B+C) - F_2(4, n-1, A, C)\right)
\right.\nonumber\\ & \left.
  + 2 A C \left(F_2(3,n-2,A-B,B+C) - F_2(3,n-2,A,C)\right)
\right.\nonumber\\ & \left.
  +C^2\left( F_2(2,n-1,A-B,B+C) - F_2(2,n-1,A,C)\right)\right).
\end{align}

{We can write our one loop functions using $F_i$'s and $F_{3,2}$ as follows; 
\begin{eqnarray}
\int \frac{dxdydzdw\delta(x+y+z+w-1)xyzw}{[x m^2_{\eta_0} +y m^2_{\chi_0} +z M^2_{N\Sigma_a}  + w M^2_{{N_L}_a}]^4}
\left[
3-\frac{4 M_{{N_L}_a}^2}{x m^2_{\eta_0} +y m^2_{\chi_0} +z M^2_{N\Sigma_a} + w M^2_{{N_L}_a}}\right]\nonumber\\
=3 \left(F_4(1, 4, m^2_{\eta_0}-M^2_{{N_L}_a}, m^2_{\chi_0}-M^2_{{N_L}_a}, M^2_{N\Sigma_a} -M^2_{{N_L}_a}, M^2_{{N_L}_a})
\right.\nonumber\\  \left.
-F_4(2, 4, m^2_{\chi_0}-M^2_{{N_L}_a}, M^2_{N\Sigma_a} -M^2_{{N_L}_a}, m^2_{\eta_0}-M^2_{{N_L}_a}, M^2_{{N_L}_a})
\right.\nonumber\\  \left.
-F_4(2, 4, m^2_{\eta_0}-M^2_{{N_L}_a}, M^2_{N\Sigma_a} -M^2_{{N_L}_a}, m^2_{\chi_0}-M^2_{{N_L}_a}, M^2_{{N_L}_a})
\right.\nonumber\\  \left.
-F_4(2, 4, m^2_{\eta_0}-M^2_{{N_L}_a}, m^2_{\chi_0}-M^2_{{N_L}_a}, M^2_{N\Sigma_a} -M^2_{{N_L}_a}, M^2_{{N_L}_a})
\right)
\nonumber\\
-4 M_{{N_L}_a}^2\left(F_5(1, 5, m^2_{\eta_0}-M^2_{{N_L}_a}, m^2_{\chi_0}-M^2_{{N_L}_a}, M^2_{N\Sigma_a} -M^2_{{N_L}_a}, M^2_{{N_L}_a})
\right.\nonumber\\  \left.
-F_4(2, 5, m^2_{\chi_0}-M^2_{{N_L}_a}, M^2_{N\Sigma_a} -M^2_{{N_L}_a}, m^2_{\eta_0}-M^2_{{N_L}_a}, M^2_{{N_L}_a})
\right.\nonumber\\  \left.
-F_4(2, 5, m^2_{\eta_0}-M^2_{{N_L}_a}, M^2_{N\Sigma_a} -M^2_{{N_L}_a}, m^2_{\chi_0}-M^2_{{N_L}_a}, M^2_{{N_L}_a})
\right.\nonumber\\  \left.
-F_4(2, 5, m^2_{\eta_0}-M^2_{{N_L}_a}, m^2_{\chi_0}-M^2_{{N_L}_a}, M^2_{N\Sigma_a} -M^2_{{N_L}_a}, M^2_{{N_L}_a})
\right). 
\end{eqnarray}
}
\end{appendix}

\end{document}